# Perturbative Yang-Mills Ground State in the Temporal Gauge

By Scott Chapman
Research Professor of Physics
Chapman University

ABSTRACT
=====
Second order corrections to the perturbative ground state wave functional and vacuum energy of a Yang-Mills theory are calculated in the temporal gauge. Using dimensional regularization, the concepts of renormalization and a running coupling constant are motivated in a spatially gauge invariant way, before the introduction of a scalar product. After definition of a scalar product, it is shown that the renormalization constants of this approach are just those of the Axial gauge.
=====

In recent years there has been a renewed interest in studying the perturbative ground state of a Yang-Mills theory in the Hamiltonian formulation. For example, second order corrections to the ground state wave functional were recently calculated in the Coulomb gauge [1]. One driver of this renewed interest was an ansatz for the Yang-Mills vacuum wave functional that reproduces the essential infrared physics of the theory [2-5].

The purpose of this paper is to derive the second order correction to the Yang-Mills perturbative ground state wave functional in the temporal gauge. In addition to being free of Gribov ambiguities, temporal gauge wave functionals are interesting since they are spatially gauge invariant and can be derived without the introduction of a scalar product. Results from calculations made in other gauges can be recovered from temporal gauge results by defining a scalar product and imposing a spatial gauge condition. In this way, results of this paper are tied back to those found in the Coulomb and Axial gauges.

Two interesting features of the perturbative ground state arise from this derivation. First, the required equation for local spatial gauge invariance directly leads to the Ward Identity requirement that two-field terms of the wave functional must only be transverse at any order of perturbation theory. Second, one notices that the concept of a running coupling constant with the famous 11/3 N beta function dependence arises naturally in a spatially gauge invariant way, even before the introduction of a scalar product.

Dimensional regularization is employed to evaluate divergent "loop" integrals embedded in the two-field terms and energy density. The Appendix includes useful formulas for dimensionally regularized divergent integrals in three dimensions. In addition to being used for the present work, the formulas can also be employed to reproduce integrals from previously published Coulomb gauge papers.

The Hamiltonian for a Yang-Mills theory is given by:

$$H = \int d^3x \left( \tfrac{1}{2} \Pi_i^a \Pi_i^a + \tfrac{1}{4} F_{ij}^a F_{ij}^a \right) \tag{1}$$

where $F_{ij}^a = \partial_i A_j^a - \partial_j A_i^a + gf^{abc} A_i^b A_j^c$ and $\Pi_i^a(\vec{x})$ is the canonical momentum of the gauge field $A_i^a(\vec{x})$. Due to working in the temporal gauge ($A_0^a = 0$), it is assumed that all fields are evaluated at some constant time t. Since all fields are always evaluated at that same time, the time dependence of fields is suppressed, and only spatial dependence is shown explicitly. Calculations are performed in the Schroedinger representation, which is defined by $\Pi_i^a(\vec{x}) = -i\,\delta/\delta A_i^a(\vec{x})$. Spatial gauge transformations are generated by the color Gauss Law operator:

$$G^a(\vec{x}) = \partial_i \Pi_i^a(\vec{x}) - gf^{abc} A_i^c(\vec{x}) \Pi_i^b(\vec{x}). \tag{2}$$

This operator commutes with the Hamiltonian and must annihilate any physical state in the temporal gauge (in order for the state to be spatially gauge invariant).

To any order $g^n$, one should be able to derive perturbative vacuum wave functionals that simultaneously satisfy both of the following equations:

$$H\Psi_n = E_n\Psi_n + O(g^{n+1}) \qquad (3)$$

$$G^a(\vec{x})\Psi_n = 0 + O(g^{n+1}) \qquad (4)$$

These equations mean that to order $g^n$, the wave functional is an eigenstate of the Hamiltonian and locally spatially gauge invariant. For the remainder of this paper, reference to gauge invariance will mean local spatial gauge invariance (excluding "large", topologically nontrival spatial gauge transformations). Since the above equations do not involve scalar products, no scalar product is needed when deriving the wave functional.

It is helpful to expand the gauge fields into complex Fourier modes

$$A_i^a(\vec{x}) = \int \frac{d^3k}{(2\pi)^{3/2}} \tilde{A}_i^a(\vec{k}) \exp(i\vec{k}\cdot\vec{x}). \qquad (5)$$

Since the gauge fields $A_i^a(\vec{x})$ are real functions, the Fourier modes must satisfy $\tilde{A}_i^{a*}(\vec{k}) = \tilde{A}_i^a(-\vec{k})$. It is convenient to separate the Fourier modes into longitudinal and transverse components:

$$\tilde{A}_{Li}^a(\vec{k}) = \hat{k}_i \hat{k}_j \tilde{A}_j^a(\vec{k}) \qquad \tilde{A}_{Ti}^a(\vec{k}) = \tilde{A}_i^a(\vec{k}) - \tilde{A}_{Li}^a(\vec{k}), \qquad (6)$$

From equation (2), it is apparent that the first term of $G^a(\vec{x})$ is only nonzero if it acts on longitudinal field components.

It is assumed that one can write the ground state wave functional in the following form:

$$\Psi_n = \exp[O_0 + gO_1 + g^2 O_2 + \ldots + g^n O_n], \qquad (7)$$

where each of the operators $O_m$ is a polynomial functional involving the gauge fields. By applying equation (4) to this expansion, one obtains the iterative equation:

$$[\partial_i \Pi_i^a(\vec{x}), O_{m+1}] = [f^{abc} A_i^c(\vec{x}) \Pi_i^b(\vec{x}), O_m]. \qquad (8)$$

This equation implies that for semi-simple Lie algebras, any 2-field terms in any of the $O_m$ must be transverse; no longitudinal 2-field terms are allowed. This can be seen as follows: For semi-simple Lie algebras where $f^{aab} = 0$, any term involving a single gauge field must have a gauge (e.g. color) index. Since the $O_m$ do not have any gauge indices, none of the $O_m$ polynomials can include a term with a single gauge field; all terms of each polynomial must have at least 2 gauge fields. Let us assume that in a 2-field term of some functional $O_{m+1}$, at least one of the two gauge fields is longitudinal. If that were the case, then the left side of (8) would not vanish. The only way to satisfy the equation would be if $O_m$ had a term with just one field, but this is not possible, as argued above. So any 2-field terms in any of the functionals $O_m$ must be comprised only of transverse fields. This applies to all orders of the perturbative expansion.

The zeroth order perturbative ground state and vacuum energy of a Yang-Mills theory are well known. They are given by

$$O_0 = -\tfrac{1}{2}\int d^3k\, k\, \tilde{A}_{Ti}^a(\vec{k}) \tilde{A}_{Ti}^a(-\vec{k}) \qquad (9)$$

$$E_0 = n_A \int \frac{d^3x\, d^3k}{(2\pi)^3} k, \qquad (10)$$

where $k = |\vec{k}|$ (a notation that will be used throughout) and $n_A$ is the number of adjoint gauge directions in the theory (e.g. $n_A = N^2 - 1$ for SU(N)).

From equation (8), the first order functional $O_1$ must involve 3 fields, at least one of which is transverse. The exact form of $O_1$ can be found by solving the part of the eigenequation (3) in which each term is proportional to g (see equation (A1) in the Appendix). One finds that

$$O_1 = -if^{abc} \int \frac{d^3k\,d^3p\,d^3p'}{(2\pi)^{3/2}} \delta^3(\vec{k}+\vec{p}+\vec{p}') k_i \tilde{A}^a_{Tj}(\vec{k}) \times$$

$$\times \left[ \frac{\tilde{A}^b_{Ti}(\vec{p}')\tilde{A}^c_{Tj}(\vec{p})}{k+p+p'} + \frac{\tilde{A}^b_{Ti}(\vec{p}')\tilde{A}^c_{Lj}(\vec{p})}{k+p'} + \frac{\tilde{A}^b_{Li}(\vec{p}')\tilde{A}^c_{Tj}(\vec{p})}{k+p} + \frac{\tilde{A}^b_{Li}(\vec{p}')\tilde{A}^c_{Lj}(\vec{p})}{k} \right] \quad (11)$$

$$E_1 = E_0 \quad (12)$$

satisfies both equations (3) and (8) to first order. As a check, the transverse-only part of $O_1$ matches the corresponding functional found when expanding the Yang-Mills ground state in the Coulomb gauge [1].

The second order functional $O_2$ includes two terms: one involving four gauge fields $O_2^{(4)}$ and one involving two gauge fields $O_2^{(2)}$.

$$O_2 = O_2^{(2)} + O_2^{(4)} \quad (13)$$

Again, by using equation (8) along with the terms in the eigenequation proportional to $g^2$, it is straightforward to solve for both $O_2$ and $E_2$. The explicit form of $O_2^{(4)}$ is somewhat complicated and is given in the Appendix since it is not directly relevant to the main discussion of the rest of this paper. As mentioned above, since gauge invariance requires any two-field term such as $O_2^{(2)}$ to be comprised of only transverse fields, its final form is simpler. However, $O_2^{(2)}$ involves divergent integrals that need to be regulated. Dimensional regularization is employed for that purpose, and calculations are performed in $d = 3 - 2\varepsilon$ dimensions. In practice, this means that momentum integrals are transformed from $d^3k \to d^d k$, having mass dimensionality given by $[d^d k] = 3 - 2\varepsilon$. In order to ensure that the argument of the exponent in (7) remains dimensionless, the dimensionalities of the bare gauge fields and bare coupling must become $[\tilde{A}^a_i(\vec{k})] = -2 + \varepsilon$ and $[g] = \varepsilon$. One may define a dimensionless bare coupling $g_0(\mu)$ through the relation $g_0 = g\mu^{-\varepsilon}$, where $\mu$ is the renormalization scale.

Solving equations (3) and (8) at second order, one finds

$$g^2 O_2^{(2)} = \tfrac{1}{2}\beta \int d^d k\, k \left( \frac{1}{\varepsilon} + c_2 - \ln\left(\frac{k^2}{\mu^2}\right)\right) \tilde{A}^a_{Ti}(\vec{k})\tilde{A}^a_{Ti}(-\vec{k}), \quad (14)$$

$$E_2 = \frac{n_A(d-1)}{2} \int \frac{d^d x\, d^d k}{(2\pi)^d} k \left[1 - \beta\left(\frac{1}{\varepsilon} + c_2 - \ln\left(\frac{k^2}{\mu^2}\right)\right)\right], \quad (15)$$

$$\beta = \frac{11 C_A g_0^2}{3(4\pi)^2}, \quad (16)$$

where $C_A$ is the adjoint Casimir operator for the group (e.g. $C_A = N$ for SU(N)), and $c_2$ is a finite numerical constant that is unimportant for the present work. Note that the famous beta factor associated with the running coupling constant of a Yang-Mills theory appears in this approach even though a scalar product has not yet been defined.

As an aside, it is worth mentioning that the types of integrals that arise in $3 - 2\varepsilon$ dimensions are quite different than the types that arise in the more common $4 - 2\varepsilon$ dimensional approaches in covariant gauges. A useful identity for $d = 3 - 2\varepsilon$ dimensional integrals is presented in (A13) of the Appendix. In addition to informing this work, this identity can also be used as an alternative way to calculate integrals presented in [1].

It is interesting to see how far one can go toward motivating a running coupling before defining a scalar product. If one defines the following constant,

$$Z_A^{-1} = 1 - \beta\left(\frac{1}{\varepsilon} + c_2\right), \quad (17)$$

then it is possible to combine $O_0$ and $O_2^{(2)}$ as follows

$$O_0 + g^2 O_2^{(2)} = Z_A^{-1} O_0 - \tfrac{1}{2}\beta \int d^d k\, k \ln\left(\frac{k^2}{\mu^2}\right) \tilde{A}_{Ti}^a(\vec{k}) \tilde{A}_{Ti}^a(-\vec{k}), \tag{18}$$

so that the divergent $1/\varepsilon$ part of $O_2^{(2)}$ has been absorbed into the constant $Z_A^{-1}$. Since $O_0$ and $O_1$ satisfy equation (8), then $Z_A^{-1} O_0$ and $Z_A^{-1} O_1$ must also satisfy equation (8). This tells us that all of the divergent $1/\varepsilon$ parts of the third order, three-field operator $O_3^{(3)}$ that have at least one longitudinal field must be incorporated into $Z_A^{-1} O_1$. If one wanted to absorb all of these divergent terms into renormalized fields and a renormalized coupling, this could be achieved by making the definitions:

$$A_i^{Ra}(\vec{k}) = Z_A^{-1/2} \tilde{A}_i^a(\vec{k}) \tag{19}$$

$$g_R = Z_\alpha^{-1/2} g_0 \tag{20}$$

with

$$Z_\alpha = 1 - \beta\left(\frac{1}{\varepsilon} + c_2\right) \tag{21}$$

Equations (20) and (21) motivate the running coupling constant in a spatially gauge invariant way and without the use of a scalar product.

It should be emphasized that without a scalar product, the running coupling is only motivated, not formally derived. To formally derive the running coupling, the renormalization procedure must be defined such that the scalar products associated with any physical observables are finite and not divergent. To make connection with this standard renormalization procedure, the scalar product will now be defined.

In the temporal gauge, the scalar product $( \, , \, )$ is defined through the introduction of a canonical gauge condition $\chi^a(A) = 0$ and its associated Faddeev Popov functional determinant $\det[G^a, \chi^b]$. For example the norm of $\Psi_2$ can be defined by:

$$(\Psi_2, \Psi_2) = \int DA_i^{Ra}(\vec{x})\, \delta[\chi^a(A)]\, \det[G^a, \chi^b]\, \Psi_2^* \Psi_2, \tag{22}$$

where $DA_i^{Ra}$ is the functional integral over renormalized gauge fields at all points of space. So, to evaluate a scalar product, one must pick a spatial gauge condition. The Axial gauge $\chi^a = A_3^a(\vec{x}) = 0$ has the advantage that the Faddeev Popov (FP) determinant is independent of the gauge fields, so it decouples from the integral. In the Axial gauge with no contributions from FP ghosts, the renormalization procedure presented above in equations (17) to (21) is valid. When expressed in terms of renormalized quantities, there will be no $1/\varepsilon$ terms in $\Psi_2$, so scalar products like that of (22) will have no divergent $1/\varepsilon$ dependence. It is no surprise then that expressions for $Z_A$ and $Z_\alpha$ of equations (17) and (21) are exactly those found in standard Axial gauge renormalization approaches [6] (except that in the present approach, the finite constant $c_2$ has been included in the renormalization constant for convenience). This agreement provides a good check of the results of (14), (15) and (16).

APPENDIX

A key assumption for this paper is that the functionals $O_m$ in the expansion (7) can be expressed as polynomials in the gauge fields. With that assumption in hand, the eigenequation (3) can be disaggregated to an infinite number of separate equations, each involving some power of $g^n$ and some definite number of gauge fields. For example, to derive $O_1$ of equation (11), one needs only to solve the part of (3) involving a single factor of g and three gauge fields. This equation is

$$\int d^3k k \tilde{A}_{Tj}^a(\vec{k}) \frac{\delta}{\delta \tilde{A}_{Tj}^a(\vec{k})} O_1 + f^{abc} \int d^3x \partial_i A_j^a(\vec{x}) A_i^b(\vec{x}) A_j^c(\vec{x}) = 0 . \tag{A1}$$

Due to its $\tilde{A}_{Tj}^a(\vec{k})$ factor, the first term above has at least one transverse field. If the second term above had a part comprised of only longitudinal fields, then it would not be possible to satisfy (A1), and one might have to question the assumption that the $O_m$ are polynomials. However, the $\partial_i A_j^a(\vec{x})$ factor in the second term means that it has at least one transverse field, so (A1) can be satisfied, and it is straightforward to see that (11) does the job. For $O_1$ to be correct, it must also satisfy (8) to first order, and it is straightforward to verify that it does.

For semi-simple Lie algebras where $f^{aab} = 0$, the one-field parts of equation (3) vanish at all orders. The zero-field first order part of equation 3 is just equation (12). One can generalize equation (12) as follows. Equation (8) implies that the functionals $O_m$ will only have terms involving an even number of gauge fields when *m* is even and an odd number of gauge fields when *m* is odd. It follows that the only nonvanishing parts of equation (3) for a given order $g^m$ will be equations with an even number of gauge fields when *m* is even and an odd number of gauge fields when *m* is odd. In particular, this implies in general that

$$E_{2n+1} - E_{2n} = 0 \tag{A2}$$

for any non-negative integer *n*.

The second order four-field part of equation (3) is given by

$$\int d^3k k \tilde{A}_{Tj}^a(\vec{k}) \frac{\delta}{\delta \tilde{A}_{Tj}^a(\vec{k})} O_2^{(4)} - \frac{1}{2}\int d^3k \frac{\delta}{\delta \tilde{A}_j^a(-\vec{k})} O_1 \frac{\delta}{\delta \tilde{A}_j^a(\vec{k})} O_1 + \frac{1}{4} f^{abc} f^{ade} \int d^3x A_i^b(\vec{x}) A_j^c(\vec{x}) A_i^d(\vec{x}) A_j^e(\vec{x}) = 0 \tag{A3}$$

If the second two terms included a part comprised solely of longitudinal fields, then due to the $\tilde{A}_{Tj}^a(\vec{k})$ factor in the first term, it would not be possible to solve (A3), and one might again question the polynomial assumption. Fortunately, the longitundinal-only parts of the second two terms of (A3) exactly cancel, so with $O_1$ in hand, it is possible to find an expression for $O_2^{(4)}$ that satisfies (A3). It should be noted that to also satisfy equation (8), $O_2^{(4)}$ must include a longitudinal-only term that is not specified by (A3), only by (8).

An explicit form of $O_2^{(4)}$ satisfying (A3) and (8) is given below, where the notations $\vec{k} = -\vec{p} - \vec{q}$, $\tilde{A}_L^b(\vec{q}) = \hat{q}_i \tilde{A}_{Li}^b(\vec{q})$ and $P_{ij}^T(\vec{k}) = \delta_{ij} - \hat{k}_i \hat{k}_j$ are employed.

$$O_2^{(4)} = -f^{abc} f^{ade} \int \frac{d^3q\, d^3p\, d^3q'\, d^3p'}{(2\pi)^3} \delta^3(\vec{p} + \vec{q} + \vec{q}' + \vec{p}')[O_2^{00} + O_2^{10} + O_2^{20} + O_2^{11} + O_2^{21} + O_2^{22}]$$

(A4)

$$O_2^{00} = \frac{1}{4k} \tilde{A}_L^b(\vec{q}) \tilde{A}_{Lj}^c(\vec{p}) \tilde{A}_L^d(\vec{q}') \tilde{A}_{Lj}^e(\vec{p}') \tag{A5}$$

$$O_2^{10} = \left\{ \frac{k_i}{kq'} - \frac{p_i'}{q'p'} \right\} \tilde{A}_{Li}^b(\vec{q}) \tilde{A}_{Lj}^c(\vec{p}) \tilde{A}_L^d(\vec{q}') \tilde{A}_{Tj}^e(\vec{p}') \tag{A6}$$

$$O_2^{20} = \left\{ -\frac{\left[\frac{1}{2}(p_k - q_k)\delta_{ij} - 2\delta_{kj}p_i\right]P_{kl}^T(\vec{k})k_m\hat{q}'_m\hat{p}'_l}{k(p+q)(k+q+p)} + \frac{\frac{1}{2}\hat{q}'_i\hat{p}'_j}{p+q} \right\} \tilde{A}_{Ti}^b(\vec{q})\tilde{A}_{Tj}^c(\vec{p})\tilde{A}_L^d(\vec{q}')\tilde{A}_L^e(\vec{p}') \qquad (A7)$$

$$O_2^{11} = \frac{1}{2}\left\{ \frac{\left[(pp' + q'_k q_k + k^2)\delta_{ij} - (q'_j q_i - q_j q'_i)\right]}{qq'(p+p')} - \frac{(k^2\delta_{ij} + q_j q'_i)}{kqq'} \right\} \tilde{A}_L^b(\vec{q})\tilde{A}_{Tj}^c(\vec{p})\tilde{A}_L^d(\vec{q}')\tilde{A}_{Ti}^e(\vec{p}') \qquad (A8)$$

$$O_2^{21} = \left\{ \frac{\frac{1}{2}(p_k - q_k)\delta_{ij}P_{km}^T(\vec{k}) - \delta_{mj}p_i + \delta_{mi}q_j}{k+q+p} + \frac{\frac{1}{2}(q-p)\delta_{ij}\hat{k}_m}{k} \right\} \tilde{A}_{Ti}^b(\vec{q})\tilde{A}_{Tj}^c(\vec{p})\tilde{A}_L^d(\vec{q}')\tilde{A}_{Tm}^e(\vec{p}') \qquad (A9)$$

$$O_2^{22} = \frac{1}{2(k+q+p)(k+q'+p')}\left\{ \frac{1}{4}\delta_{jk}\delta_{lm}\left[ \frac{(q_i-p_i)(q'_i-p'_i) - (q-p)(q'-p')}{(q+p+q'+p')} - \frac{(q-p)(q'-p')}{k} \right] \right.$$
$$-\delta_{jk}\frac{(p_m - q_m)k_l}{(q+p+q'+p')} + \delta_{lm}\frac{(p'_k - q'_k)k_j}{(q+p+q'+p')}$$
$$\left. + \delta_{km}\left[ -\frac{4k_j k_l}{(q+p+q'+p')} + \frac{\delta_{jl}(k+q+p)(k+q'+p')}{2(q+p+q'+p')} \right] \right\} \tilde{A}_{Tj}^b(\vec{q})\tilde{A}_{Tk}^c(\vec{p})\tilde{A}_{Tl}^d(\vec{q}')\tilde{A}_{Tm}^e(\vec{p}') \qquad (A10)$$

The second order two-field part of equation (3) is given by

$$\int d^d k k \tilde{A}_{Tj}^a(\vec{k}) \frac{\delta}{\delta \tilde{A}_{Tj}^a(\vec{k})} O_2^{(2)} - \frac{1}{2}\int d^d q \frac{\delta}{\delta \tilde{A}_j^a(-\vec{q})} \frac{\delta}{\delta \tilde{A}_j^a(\vec{q})} O_2^{(4)} = 0 . \qquad (A11)$$

Just as in similar equations discussed earlier, if the second term above contained a purely longitudinal part, then it would not be possible to solve the equation. Fortunately, it does not. So with $O_2^{(4)}$ in hand, one can calculate $O_2^{(2)}$. As mentioned in the main body of the paper, the calculation is divergent, and dimensional regularization is employed with $d = 3 - 2\varepsilon$. Taking $\vec{q}$ as the loop momentum and $\vec{p}$ as the momentum of the remaining fields, the second term of (A11) depends on $|\vec{q}|$ and $|\vec{q}+\vec{p}|$ in combinations like $|\vec{q}+\vec{p}|+|\vec{q}|$ and $|\vec{q}+\vec{p}|-|\vec{q}|$. These are different combinations than those that arise in 4-dimensional calculations, but they can be evaluated using the identities given below.

Define the following quantities:

$$K = |\vec{q}+\vec{p}| + |\vec{q}| \qquad \text{and} \qquad Q = |\vec{q}+\vec{p}| - |\vec{q}| . \qquad (A12)$$

For small $\varepsilon$, the following identity holds

$$\int \frac{d^d q}{|\vec{q}||\vec{q}+\vec{p}|} f(K,Q,p) = \pi\left(\frac{\pi}{4p^2}\right)^{-\varepsilon} \Gamma(1+\varepsilon)\int_p^\infty dK(K^2 - p^2)^{-\varepsilon}\int_{-p}^p dQ(p^2 - Q^2)^{-\varepsilon} f(K,Q,p) + O(\varepsilon). \qquad (A13)$$

This identity can be derived in the following way: First, on the left side insert a factor of one in the form of integrals over delta functions:

$$1 = 4\int_0^\infty dk_0 k_0 dq_0 q_0 \frac{dwdv}{(2\pi)^2} \exp\left[i(k_0^2 - (q_i + p_i)^2)w + i(q_0^2 - q_i^2)v\right] \qquad (A14)$$

Next, replace factors of $|\vec{q}|$ and $|\vec{q}+\vec{p}|$ with $q_0$ and $k_0$, and let $v \to v - w$. Then, using known integral expressions for $\cos(ax^2)$ and $\sin(ax^2)$, perform the integrals over $d^d q$ and $dw$. Next, perform the $dv$ integral by using known expressions for the integrals of $v^{-\beta}\cos(av)$ and $v^{-\beta}\sin(av)$. Rewrite the remaining integrals over $q_0$ and $k_0$ to be over $k_0 + q_0$ and $k_0 - q_0$, which are equivalent to $K$ and $Q$ by the delta functions. After this, one obtains the right hand side of equation (A13).

In (A13), it is apparent that any functions $f(K,Q,p)$ that are odd in Q will vanish in the above integral. The Q integrals are convergent due to their finite limits. The K integrals, however can be divergent. In particular, for logarithmically divergent integrals, the following K integral is useful:

$$2\pi^{-\varepsilon}\mu^{2\varepsilon}\Gamma(1+\varepsilon)\int_{p}^{\infty}\frac{dK}{(K+p)}(K^2-p^2)^{-\varepsilon}=\left\{\frac{1}{\varepsilon}-\ln\left(\frac{p^2}{\mu^2}\right)-\ln(4\pi)+1-\gamma\right\}+O(\varepsilon) \qquad (A15)$$

For linearly and quadratically divergent integrals, the following identities are useful

$$2\pi^{-\varepsilon}\mu^{2\varepsilon}\Gamma(1+\varepsilon)\int_{p}^{\infty}dK(K^2-p^2)^{-\varepsilon}=-2p+O(\varepsilon) \qquad (A16)$$

$$2\pi^{-\varepsilon}\mu^{2\varepsilon}\Gamma(1+\varepsilon)\int_{p}^{\infty}dKK(K^2-p^2)^{-\varepsilon}=0+O(\varepsilon), \qquad (A17)$$

where the usual approach is employed of first taking d to be small enough for the integrals to be convergent, then analytically continuing d to the value $d=3-2\varepsilon$ at the end of the calculation. In addition to determining the form of $O_2^{(2)}$ in (14), these identities can also be used to reproduce the integrals $I_c$ and $I_g$ found in [1].

The second order no-field part of equation (3) is given by

$$-\tfrac{1}{2}\int d^d k \frac{\delta}{\delta\tilde{A}_j^a(-\vec{k})}\frac{\delta}{\delta\tilde{A}_j^a(\vec{k})}g^2 O_2^{(2)} = E_2 - E_0 \qquad (A18)$$

The form of the second order vacuum energy shown in (15) follows directly from this equation.